\documentclass[preprint2]{aastex}
\shorttitle{HDE 245770 and X Per}
\shortauthors{Grundstrom et al.}
\begin{document}

\received{2006 October 27}

\title{Joint H$\alpha$ and X-Ray Observations of Massive X-Ray Binaries. III.
The Be X-ray Binaries HDE 245770 = A 0535+26 and X Persei} 

\author{E. D. Grundstrom\altaffilmark{1}, T. S. Boyajian\altaffilmark{1}, 
 C. Finch, D. R. Gies\altaffilmark{1}, W. Huang\altaffilmark{1,2}, \\
 M. V. McSwain\altaffilmark{1,3,4}, 
 D. P. O'Brien, R. L. Riddle\altaffilmark{1,5}, M. L. Trippe, 
 S. J. Williams\altaffilmark{1}, \\ D. W. Wingert\altaffilmark{1}, and R. A. Zaballa}
\affil{Center for High Angular Resolution Astronomy and \\
 Department of Physics and Astronomy,\\
 Georgia State University, P. O. Box 4106, Atlanta, GA 30302-4106; \\
 erika@chara.gsu.edu, tabetha@chara.gsu.edu, 
 finch@chara.gsu.edu, gies@chara.gsu.edu, wenjin@astro.caltech.edu,
 mcswain@astro.yale.edu, obrien@chara.gsu.edu, riddle@tmt.org, trippe@chara.gsu.edu, 
 swilliams@chara.gsu.edu, wingert@chara.gsu.edu, rzaballa1@student.gsu.edu}

\altaffiltext{1}{Visiting Astronomer, Kitt Peak National Observatory,
National Optical Astronomy Observatory, operated by the Association
of Universities for Research in Astronomy, Inc., under contract with
the National Science Foundation.}
\altaffiltext{2}{Current address: Department of Astronomy, 
 California Institute of Technology, MC 105-24, 
 Pasadena, CA 91125}
\altaffiltext{3}{Current Address: Astronomy Department,
Yale University, New Haven, CT 06520-8101}
\altaffiltext{4}{NSF Astronomy and Astrophysics Postdoctoral Fellow}
\altaffiltext{5}{Current Address: Thirty Meter Telescope,
 2632 E. Washington Blvd., Pasadena, CA 91107}

\slugcomment{Submitted to ApJ.}
\paperid{70681}


\begin{abstract}

We present results from an H$\alpha$ monitoring campaign 
of the Be X-ray binary systems HDE~245770 = A~0535+26 and X~Per.
We use the H$\alpha$ equivalent widths together with adopted 
values of the Be star effective temperature, disk inclination, 
and disk outer boundary to determine the half-maximum emission 
radius of the disk as a function of time. 
The observations of HDE~245770 document the rapid spectral variability
that apparently accompanied the regeneration of a new circumstellar 
disk.  This disk grew rapidly during the years 1998 -- 2000, 
but then slowed in growth in subsequent years. 
The outer disk radius is probably truncated by resonances between 
the disk gas and neutron star orbital periods.  Two recent 
X-ray outbursts appear to coincide with the largest 
disk half-maximum emission radius 
attained over the last decade.  Our observations of X~Per 
indicate that its circumstellar disk has recently 
grown to near record proportions, and concurrently the 
system has dramatically increased in X-ray flux, 
presumably the result of enhanced mass accretion from the 
disk.  We find that the H$\alpha$ half-maximum emission radius
of the disk surrounding X~Per reached a size 
about six times larger than the stellar radius, 
a value, however, that is well below the minimum separation between 
the Be star and neutron star.  We suggest that spiral arms 
excited by tidal interaction at periastron may help lift 
disk gas out to radii where accretion by the neutron star 
companion becomes more effective. 

\end{abstract}



\setcounter{footnote}{5}

\section{Introduction}                              

Be X-ray binaries (BeXRBs) consist of rapidly rotating B-stars 
with neutron star companions \citep{coe00}. 
The B-star primaries lose mass into an outflowing circumstellar 
disk, and if the disk reaches a radius comparable to the periastron
separation, then disk gas accreted by the neutron star can 
power a significant (and often transient) X-ray source \citep{oka01,oka02}.
The disk gas is detected through observations of emission lines, 
an optical/infrared excess, and a net linear polarization caused by 
starlight scattered in the disk \citep{por03}.   Be star disks are 
inherently time variable and can develop and disappear on timescales 
of years to decades \citep{und82,hub98}.  Thus, we expect that the 
X-ray accretion fluxes will vary on similar timescales.  
Here we present a joint study of the disk H$\alpha$ emission flux 
and X-ray flux variations in two of the best studied BeXRBs, 
HDE~245770 = A~0535+26 and X~Persei.  
We discuss similar co-variations of H$\alpha$ emission and X-ray flux 
in companion papers on the black hole binary Cyg~X-1 = HDE~226868 \citep{gie03},
the supergiant binary LS~I~+65~010 \citep{gr07a}, 
and the BeXRB microquasar LS~I~+61~303 \citep{gr07b}.

The remarkable Be star HDE~245770 (and X-ray counterpart 
A~0535+26) is one of the prototype objects of the BeXRB class. 
The primary star of HDE~245770 has a classification of 
O9.7~IIIe \citep{gio92} or B0~IIIe \citep{ste98}, but the 
luminosity class is still somewhat controversial since 
many of the luminosity criteria are based on spectral lines 
that are affected by disk emission \citep{wan98}.  
The orbital period from peaks in the X-ray flux is 
approximately 110~d \citep{hut84,coe06}, and a comparison 
of the times of maxima between the first and most recent 
epochs suggests a period of $P=109.96\pm0.05$~d. 
The only reliable estimate of the other orbital elements 
comes from X-ray timing of the pulsar spin period during 
three outbursts in 1993 \citep{fin94,bil97}, and this solution 
indicates that the orbital eccentricity is $e=0.47\pm 0.02$.
The radial velocity curve of the Be star has an amplitude 
that is smaller than or comparable to the velocity fluctuations
introduced by emission contamination and other factors, 
so only upper limits are available for the Be star's 
semiamplitude and the system mass ratio, $M_2/M_1<0.12$  
\citep{wan98}.  The star has a long and rich history of X-ray, 
optical, and infrared observations \citep*{mot91,gio92,cla98,neg98,hai04,coe06}. 
\citet{lyu00} present a remarkable record of the light curve
variations back to the year 1898, and they show that after a 
quiescent stage in the first part of the century the star 
brightened by some $40\%$ in the early 1970s.  The circumstellar 
disk that appeared then provided the gas to power the X-ray source 
at the time coincident with the dawn of X-ray astronomy.  

The emission from the disk of HDE~245770 in the form of optical and infrared 
continuum light and Balmer emission lines displays large temporal 
variations in the record since the 1970s \citep{cla98,lyu00,hai04}. 
The dominant timescales of variability are $\sim1500$~d and 103~d, 
where the latter is the beat period of the long and orbital 
periods \citep{hai04}.  \citet{hai04} argue that the disk IR flux 
tends to hover around three different brightness levels, 
and they suggest that these correspond to distinct disk radii 
that are defined by resonances between the disk gas and neutron
star orbital periods.  \citet{oka01} developed a model to 
predict the resonant disk truncation radii in BeXRBs, and 
\citet{hai04} found that the model predictions are consistent 
with the IR magnitude jumps.  \citet{coe06} extended this 
analysis with additional $JHK$ photometry and measurements of
the H$\alpha$ emission line.  The H$\alpha$ emission variations
\citep{cla98,hai99,lyu00,hai04,coe06} indicate that the disk almost 
totally disappeared in 1998 but rebounded to its former strength
over the next few years. 

The binary system X~Per (HD24534; B0~Ve, \citealt{lyu97}) is the 
brightest and perhaps most famous member of the BeXRB class. 
X~Per has displayed time variable optical emission lines 
since the earliest photographic spectrograms were made at the 
beginning of the last century \citep{cow72}.   The long term 
changes in the spectral appearance are reviewed by \citet{roc93} 
and \citet{cla01}.   In a seminal paper, \citet{cla01}
describe the photometric and spectroscopic variations observed 
over the period 1987 through 2001, and during this time the star
made a remarkable transformation from Be to B and back again to Be. 
\citet{cla01} discuss how these variations are related to the 
structural properties of the circumstellar disk. 
The orbital elements were determined by 
\citet{del01} by careful timing observations of the X-ray pulsar 
(pulse period $\approx 835$~s) using the {\it Rossi X-Ray Timing Explorer 
(RXTE)} satellite.  The orbital period is $P=250.3$~d and the
orbital eccentricity is $e=0.11$.  

Both HDE~245770 and X~Per are among some hundred objects regularly observed with the 
{\it RXTE} All-Sky Monitor (ASM) instrument \citep{lev96}, and here 
we address the issue of how their disk size variations are related to the 
observed X-ray fluxes.  We present our recent observations of their 
H$\alpha$ emission lines in \S2.  We then show in \S3 how the 
H$\alpha$ emission strength is related to the disk radius, and 
we use these relations in \S4 to document how the disk radius 
variations relate to the $V$-band and X-ray light curves.
The disk radii appear to attain limits that are probably related
to resonant truncation radii \citep{oka01}, and we discuss in \S5 
how such limits affect the mass transfer process in these binaries. 
An appendix to the paper offers a reassessment of 
the orbital motion of X~Per based upon radial velocity measurements  
of UV spectra from the {\it International Ultraviolet 
Explorer (IUE)} satellite, from which we derive limits on 
the binary's mass ratio and inclination.


\section{Observations and H$\alpha$ Variations}     

We observed both HDE~245770 and X~Per between 1998 August and 
2000 December with the Kitt Peak National Observatory 0.9~m 
coud\'{e} feed telescope.  We used two spectrograph 
arrangements to record the red spectral region around H$\alpha$
with resolving powers of $R=\lambda / \triangle \lambda = 4100$ 
and 9500.  The details about the spectra and their reduction  
for these runs are given in a companion paper \citep{gr07a}.  
In addition to these primary runs, we obtained several more 
spectra during auxiliary runs in 2004 October and 2006 October 
(using grating B in second order for a resolving power $R=9500$
over the range 6470 -- 7140 \AA ~and 6433 -- 7143 \AA , respectively).  
All the spectra record the H$\alpha$ emission line and 
the \ion{He}{1} $\lambda 6678$ feature.   

The general trends in the spectra are illustrated in Figures 
1 -- 3.   We show in Figure~1 the set of spectra of HDE~245770 
from the 1998 August -- September run that coincided with the 
time of disk loss.  The spectra are arranged with their continua 
set at the time of observation and scaled so that 1~d along the 
$y$-axis corresponds to $10\%$ of the continuum flux.  We also 
show a model spectrum at the bottom of the figure that is derived 
from the grid of non-local thermodynamic equilibrium and line-blanketed 
model atmospheres calculated by \citet{lan03}.  This synthetic 
spectrum is based on the parameters of effective temperature 
$T_{\rm eff} = 28000$~K, gravity $\log g = 3.3$, and solar abundances
\citep{gio92}.   The spectrum was broadened by a simple 
convolution with a rotational broadening function using a 
linear limb darkening coefficient of 0.24 \citep{wad85} and a 
projected rotational velocity of $V\sin i = 230$ km~s$^{-1}$
\citep{gio92,wan98,hai04}.  We see that there is evidence of 
residual emission components in both the H$\alpha$ and 
\ion{He}{1} $\lambda 6678$ absorption lines at this time. 
The profiles also display significant night-to-night or faster 
variations.  For example, there are well defined structures 
within the core of \ion{He}{1} $\lambda 6678$ that may result 
from photospheric nonradial pulsations (see the case of the
similar star $\zeta$~Oph; \citealt{kam97}), and there are 
variations in the relative strengths of the violet and red 
emission peaks in H$\alpha$.  These suggest that mass loss 
processes leading to the development of the new disk were 
already underway at this time.  Note that the model indicates 
that a line blend should be present for the \ion{C}{2} 
$\lambda 6578,6582$ doublet, which appears to be absent 
from the HDE~245770 spectra.  These lines weaken at higher 
temperature, but an increase in temperature would result in a
model \ion{He}{1} $\lambda 6678$ profile weaker than observed. 
The absence of the \ion{C}{2} features (and the weakness of 
the \ion{C}{4} profiles in the UV; \citealt{wan98}) may point to 
a carbon deficiency caused by the presence of CNO-processed 
gas in the atmosphere of HDE~245770. 

\begin{figure}
\begin{center}
{\includegraphics[angle=90,height=12cm]{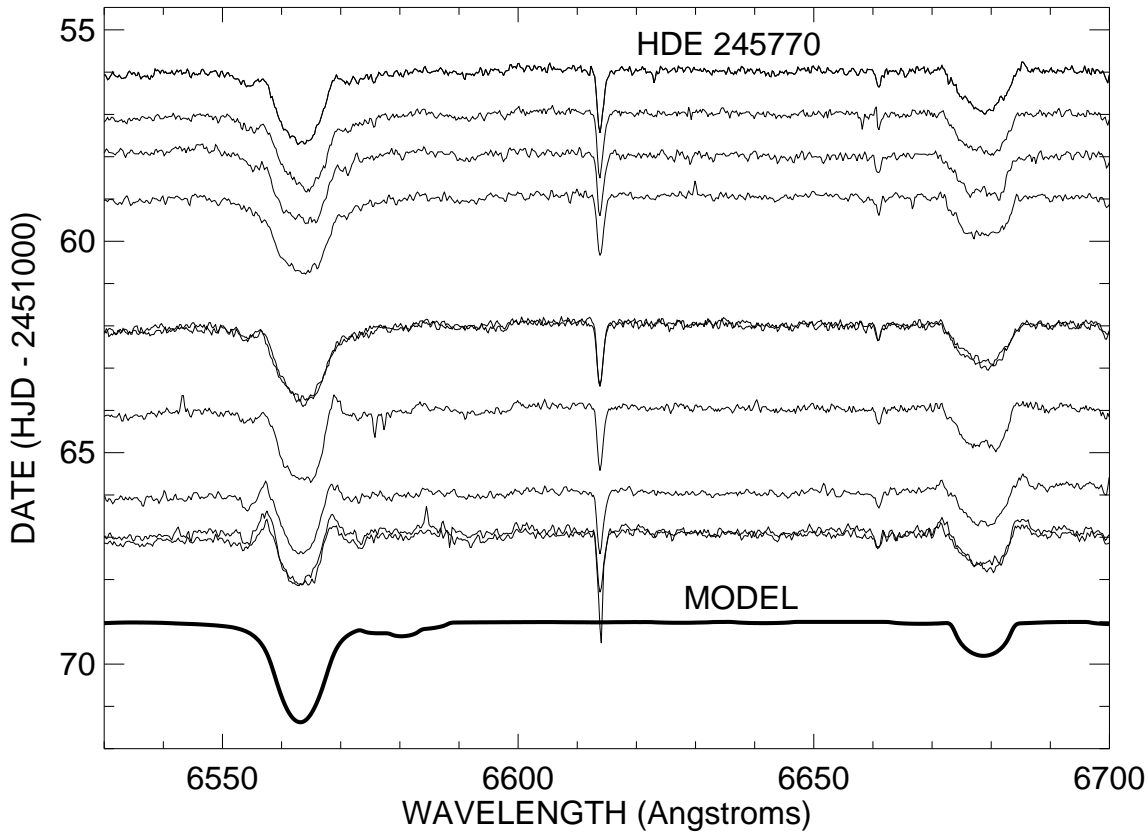}}
\end{center}
\caption{A sequence of red spectra of HDE~245770 from 1998 when the 
the disk had almost disappeared.  Each spectrum has its continuum 
aligned with the heliocentric Julian date of observation, and each 
is scaled in flux so that $10\%$ of the continuum equals 1~d of time. 
The bottom synthetic spectrum represents a model with 
$T_{\rm eff}=28000$~K, $\log g = 3.3$, and $V\sin i = 230$ km~s$^{-1}$. 
The features present include the stellar H$\alpha$ $\lambda 6563$ and 
\ion{He}{1} $\lambda 6678$ lines and the interstellar features at
6613 \AA ~and 6660 \AA .}
\label{fig1}
\end{figure}

\placefigure{fig1}     

The disk emission strength of HDE~245770 had increased dramatically by the 
time of the next runs in 1999.  We show in Figure~2 the average spectrum 
from each of the runs with the continuum levels offset for clarity. 
The growth in the H$\alpha$ emission is accompanied by the appearance
of emission in \ion{He}{1} $\lambda 6678$ that almost totally 
obscures the profile during the runs in 2000.   Note how the 
separation between the violet and red peaks of H$\alpha$ decreases 
as the emission strengthens. 

\begin{figure}
\begin{center}
{\includegraphics[angle=90,height=12cm]{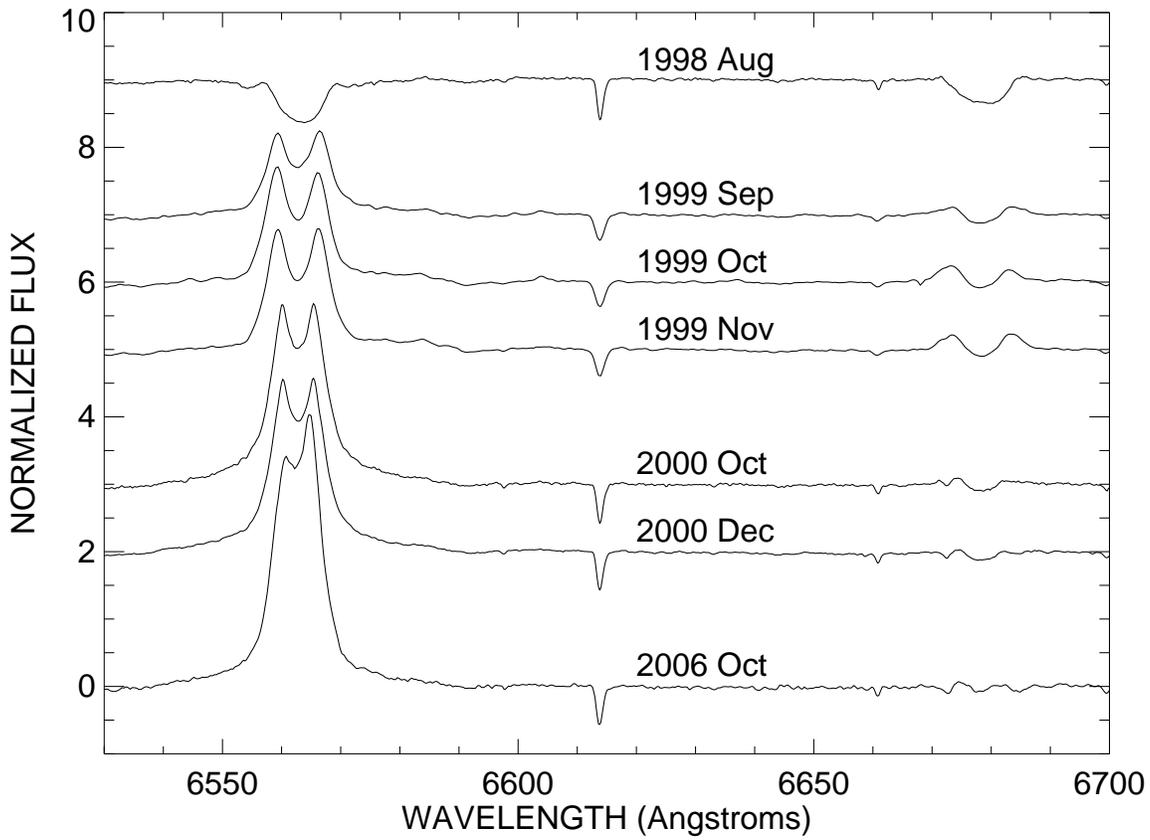}}
\end{center}
\caption{The average spectra of HDE~245770 from each of the runs 
plotted with their continua offset for clarity.  
Spectral intensity is scaled so that $100\%$ of the continuum 
equals four units of normalized flux.}
\label{fig2}
\end{figure}

\placefigure{fig2}     

We show in Figure~3 the average spectrum of X~Per from each of 
the runs, and we see that H$\alpha$ and \ion{He}{1} $\lambda 6678$
were emission lines during this time.  
The overall morphology and strength of the lines agree well with 
the more extended time series presented by \citet{cla01}. 
Inspection of Figure~7 in \citet{cla01} shows that our 
observations were made while the disk was active and generally 
increasing in strength. 

\begin{figure}
\begin{center}
{\includegraphics[angle=90,height=12cm]{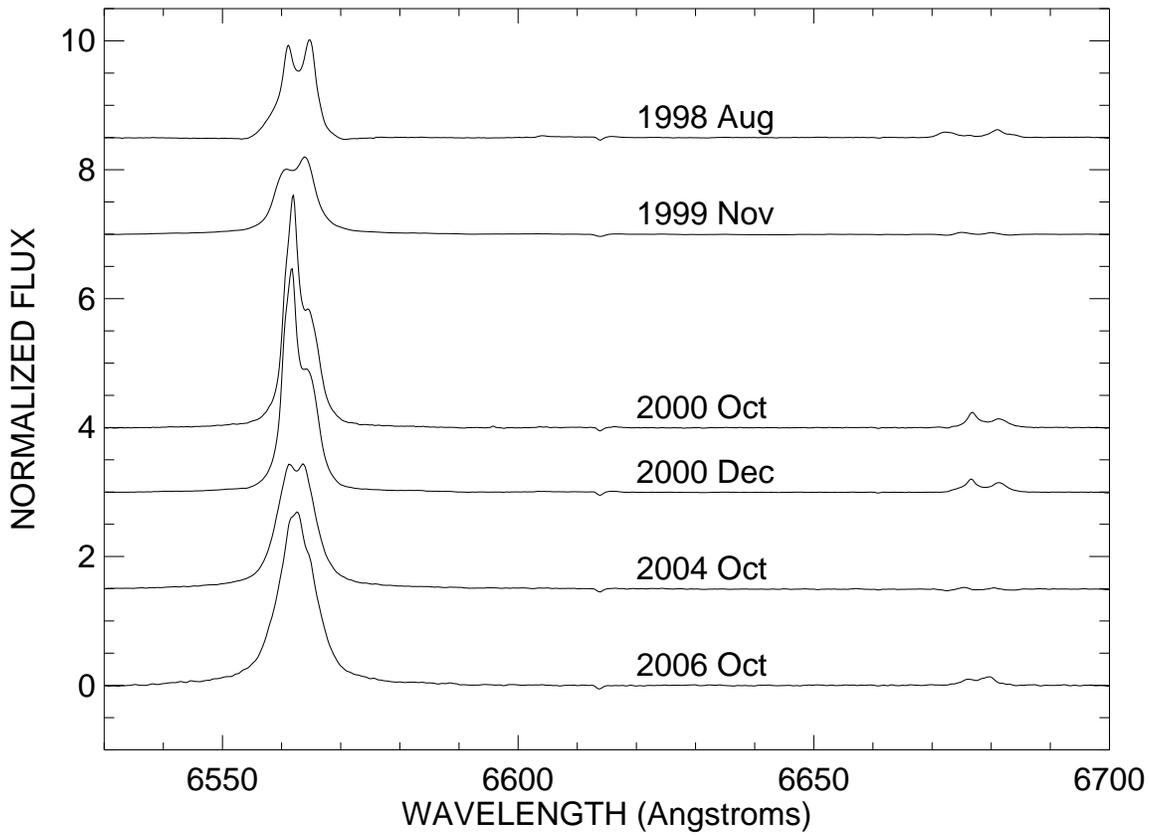}}
\end{center}
\caption{The mean spectra of X~Per from each run (separated 
in rectified intensity for clarity).
Spectral intensity is scaled so that $100\%$ of the continuum 
equals one unit of normalized flux.}
\label{fig3}
\end{figure}

\placefigure{fig3}     

We made a number of measurements to characterize the profile
variations of H$\alpha$ for both stars, and these are summarized 
in Table~1 (given in full in the electronic version). 
Column~1 lists the target name, 
column~2 gives the heliocentric Julian date of mid-exposure,  
and column~3 shows the corresponding orbital phase 
(from the ephemeris of \citealt{fin94} for HDE~245770 and 
from \citealt{del01} for X~Per).  Column~4 lists the equivalent 
width determined by a numerical integration of
the line flux over the range 6536 -- 6590 \AA . 
We measured the radial velocity of the wings based upon a bisector position
determined using the method of \citet*{sha86}.
This method samples the line wings using oppositely signed 
Gaussian functions and determines the mid-point position 
between the wings by cross-correlating these Gaussians with the profile. 
We used Gaussian functions with FWHM = 200 (100) km~s$^{-1}$ at sample 
positions in the wings of $\pm320 ~(\pm190)$ km~s$^{-1}$ for HDE~245770
(X~Per), and these radial velocities are
given in column 5 of Table~1.   \citet{zam99} advocated
making fits of double-peaked H$\alpha$ profiles using
Gaussian functions to match the violet $V$ and red $R$ peaks,
and we have followed their approach here, although we 
caution that such a functional fit is poor in many instances.
These double-Gaussian fits were restricted to the inner
part of the profile ($|\triangle \lambda | < 6$ \AA ~for
HDE~245770 and $< 2-4$ \AA ~for X~Per)
since the wings are much more extended that those of
Gaussian functions.   The remaining columns in Table~1
list the parameters for these fits:
radial velocity of the $V$ peak (col.~6),
radial velocity of the $R$ peak (col.~7),
ratio of the equivalent widths of the $V$ and $R$ components (col.~8),
$V/R$ peak intensity ratio (col.~9),
FWHM for the $V$ peak (col.~10), and
FWHM for the $R$ peak (col.~11).  
Note that H$\alpha$ has only one peak in the 
2006 observations of X~Per, so we fit a single Gaussian to 
the profiles of these three spectra. 

\placetable{tab1}      

We find no compelling evidence that these fitting parameters 
vary with orbital phase, although we caution that our orbital 
phase coverage is incomplete.  In particular, the radial velocity 
measurements for the H$\alpha$ wings and emission peaks do not 
appear follow the orbital motion expected for the Be star 
(see \citealt{wan98} for HDE~245770 and the Appendix for X~Per). 
We suspect that these velocity measurements are dominated 
by fluctuations related to the azimuthal distribution and 
non-Keplerian motion of the disk gas, and these variations 
confound attempts to measure the much smaller orbital motion in 
our data sets.  The best strategy for measuring the 
orbital radial velocities of these Be stars may be to wait until 
an episode of disk loss occurs and then embark on a program of 
high dispersion, high S/N spectroscopy of emission-free lines
in the blue part of the spectrum.


\section{Disk Radius and H$\alpha$ Strength}        

The H$\alpha$ emission forms primarily in the disk surrounding 
the Be star, and the total flux of the feature (measured as 
the line equivalent width) is closely related to the size of
the disk.  In a recent paper \citep{grg06} we present calculations
of the disk size and emission strength for an extensive range
of stellar and disk properties.  The Be disk models are based 
upon a simple parameterization of disk properties outlined 
by \citet{hum00}, from which we derive a synthetic H$\alpha$ line 
profile and the angular distribution on the sky of the 
wavelength integrated H$\alpha$ flux.   The model predictions 
about the emission strength and angular size of the disks 
agree with observational results for bright Be stars whose 
disks have been resolved through optical long baseline interferometry. 

The disk model assumes an axisymmetric gas structure with the 
gas density distribution set by a power law $R^{-m}$ (where $R$ is 
the distance from the rotational axis), an exponential decline above 
and below the disk, and a base density at the star's surface. 
The disk surface brightness is set by its temperature, and we 
assume an isothermal disk with a gas temperature of $0.6 T_{\rm eff}$
where $T_{\rm eff}$ is the effective temperature of the Be star. 
The spatial flux distribution is integrated along segments 
normal to the major axis of the projected disk to obtain 
a collapsed emission sum as a function of position along the major axis. 
We adopted a working definition of disk radius as the distance along the major axis 
from the star where the integrated H$\alpha$ emission intensity of the disk   
declines to half of the peak value found at a position immediately adjacent 
to the photosphere (the half-width at half-maximum intensity radius or HWHM radius for short).  

The functional relationship between H$\alpha$ equivalent width and
disk HWHM radius for a specific Be star depends upon the star's $T_{\rm eff}$, 
the inclination of the disk normal to the line of sight, and the 
adopted outer boundary for the disk radius \citep{grg06}. 
We list in Table~2 the values of these parameters that we adopt
for HDE~245770 and X~Per.  The listed radii and masses are estimates from 
\citet{oka01} and \citet{lyu97} and the stellar effective temperatures 
are from \citet{gio92} and \citet{lyu97} for HDE~245770 and X~Per, respectively.  
We assume that the disk is co-planar with the orbit, and we adopt 
the mid-range estimates of inclination from \citet{wan98} and the Appendix
for HDE~245770 and X~Per, respectively.  We set the outer boundary 
for the model disk equal to the Be star's Roche radius at apastron. 
The results are generally insensitive to this last assumption, since the 
H$\alpha$ flux is small from the outer, optically thin portions of the disk.
The predicted relationships between equivalent width $W_\lambda[{\rm H}\alpha]$ and 
the ratio of the disk HWHM radius to the stellar radius $R_d/R_s$ 
are shown in Figure~4 for a large range in disk base density.  The model equivalent 
width is referenced to the photospheric flux, and if the disk contributes
a continuum flux fraction of $\epsilon = F^d_\lambda / F^s_\lambda$ then 
the observed equivalent width must be prorated by a factor of $(1+\epsilon)$
in order to find the ratio of the disk to star radius $R_d/R_s$ from Figure~4. 

\placetable{tab2}      
 
\begin{figure}
\begin{center}
{\includegraphics[angle=90,height=12cm]{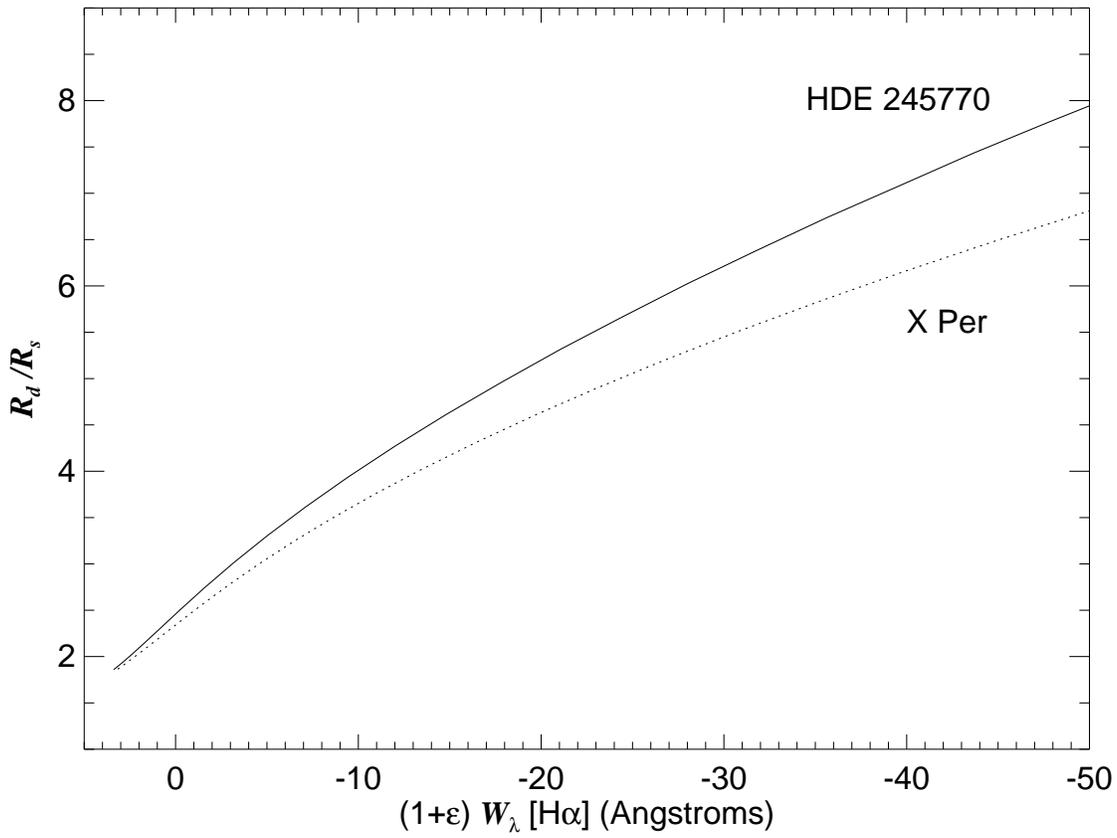}}
\end{center}
\caption{The model relation between rescaled H$\alpha$ emission equivalent
width and the ratio of disk radius (defined as the half maximum intensity 
radius of the integrated H$\alpha$ flux) to the stellar radius.}
\label{fig4}
\end{figure}

\placefigure{fig4}     


There are several features of these simple models that need to be considered 
in the derivation of a disk HWHM radius from the H$\alpha$ equivalent
width.  First, if the disk temperature differs from our assumed value of 
$0.6 T_{\rm eff}$, then the predicted H$\alpha$ equivalent width will change by
a factor of approximately $T_{\rm disk} / (0.6 T_{\rm eff})$ for a given 
disk radius.  This is due to the fact the disk gas source function varies almost 
linearly with temperature for H$\alpha$ in hot Be stars like these two targets. 
Second, the equatorial gas density is assumed to follow a power law 
$R^{-m}$ with $m=3.0$, the value often derived in studies of the IR flux
excess of Be stars.   The numerical relationship between $W_\lambda[{\rm H}\alpha]$
and $R_d/R_s$ is relatively insensitive to the selected value of $m$ \citep{grg06},
although the base density for any position along the curve depends critically on $m$.
We caution, however, that the actual disk density properties may vary 
significantly from a power law in these BeXRBs.  For example, if the star 
experiences an episode of increased disk mass loss, a density enhancement 
may appear to propagate outwards over time and produce increased emission strength
when the enhancement reaches the radius of the optically thick/thin boundary.  
Using the disk radius derived from the relation in Figure~4, we would correctly 
determine that the HWHM radius had increased but we would err if we assumed
that the emission strengthening results from an increase in density at all radii.  

Finally, it is possible the disk may be truncated by 
tidal or resonant effects at a radius smaller than our assumed outer boundary.  
If the disk density drops to zero in the outer optically thin zone, 
then only a small part of the total emission is lost and we would 
only slightly overestimate the emission flux for a given HWHM radius.   
Thus, the revised curves would appear slightly above those shown in 
Figure~4 (see an example of the variation with outer boundary in Fig.~1 of 
\citealt{grg06}) and our derived HWHM radii will be slightly smaller than 
the actual values.  On the other hand, if the disk density vanishes 
at a radius where the gas is optically thick, then the revised curves
will reach an asymptotic limit slightly below the cut-off radius.
The resulting functions would fall below the standard curves 
at the high emission end in Figure~4, and we would then tend to overestimate 
the HWHM radii using the standard relations.  The former
case probably applies to the two targets we analyze here, since the HWHM 
radii we find are generally small compared to the suspected truncation radii
(we discuss this further in \S4).
Thus, the HWHM radii measurements we derive may slightly underestimate 
the actual values.  Our analysis is insufficient to detect the presence 
of a truncation radius, but if one exists, its radius will be larger 
than the HWHM radius. 


\section{Disk Growth and X-ray Accretion Flux}      

The mass transfer rate and subsequent accretion-driven X-ray flux 
will clearly depend on the changes in the radius of the Be star's disk  
that we can track through the variations in the H$\alpha$ 
equivalent widths.  Here we consider the long term variations 
in the Be star's disk as observed in the H$\alpha$ emission 
flux and in the disk continuum flux (as seen in the observed 
$V$-band light curve).   We then compare the derived disk 
radius variations with those observed in the X-ray light 
curves from the {\it RTXE}/ASM instrument \citep{lev96}.

\subsection{HDE 245770}

\citet{lyu00} note that the $V$-band continuum of HDE~245770 may be much brighter 
than the stellar continuum alone (due to continuum light from the disk), 
and consequently the observed equivalent width referred to the 
combined continuum flux may underestimate the absolute H$\alpha$ 
emission flux.  In order to compare the H$\alpha$ emission flux to 
a constant stellar continuum, we need to rescale the equivalent width 
by the factor $1+\epsilon$ where $\epsilon$ gives the ratio 
of the monochromatic disk-to-stellar continuum flux at wavelengths
near H$\alpha$.   Following \citet{lyu00}, we can estimate the 
rescaled flux by comparing the $V$ magnitude at the time of the 
H$\alpha$ measurement with the magnitude of the star alone, $V=9.50$, by
\begin{equation}
(1+\epsilon) ~W_\lambda = 10^{-0.4(V-9.50)} ~W_\lambda.
\end{equation}
\citet{lyu00} present $V$ magnitudes for dates contemporaneous 
with the H$\alpha$ measurements through 1998, and we have used these to make 
the small corrections from equation (1).  However, there 
are no published estimates of $V$ for the succeeding years, so  
instead we relied on surrogate measurements of the $J$ 
magnitude from \citet{coe06} in order to make these corrections. 
\citet{hai04} (see their Fig.~1) show how the infrared magnitude variations 
track those observed in the $V$-band for HDE~245770, and we used 
a series of photometric $V$ \citep{lyu00} and $J$ observations \citep{coe06}
over a common time span to find a relation between these for HDE~245770,
\begin{equation}
(V-9.50) = 0.32 (J-8.67),
\end{equation}
which has an empirical scatter of $\pm 0.03$ mag.  We used the observed 
$J$ magnitudes from \citet{coe06} with equation (2) to find $V$ and 
transform the equivalent widths from 1999 onwards to an absolute scale.
We then used Figure~4 to find estimates of the disk HWHM emission
radius for the times of the H$\alpha$ observations.  

The observed light curve and derived disk HWHM radii are plotted 
in the middle and lower panels of Figure~5.  The disk HWHM radii 
are derived from our $W_\lambda$ data together
with like measurements from \citet{gio99},
\citet{lyu00}, \citet{pia00}, \citet{hai04}, and \citet{coe06}. 
We see that the disk almost disappeared in 1998 and then started to 
grow rapidly in radius over the next two years. 
The expansion rate slowed considerably by 2001 at a radius of 
$R_d/R_S \approx 4$.  However, subsequent observations
show that the disk continued to grow slowly and reached a maximum 
radius of $R_d/R_S \approx 5$ by 2005. 

\begin{figure}
\plotone{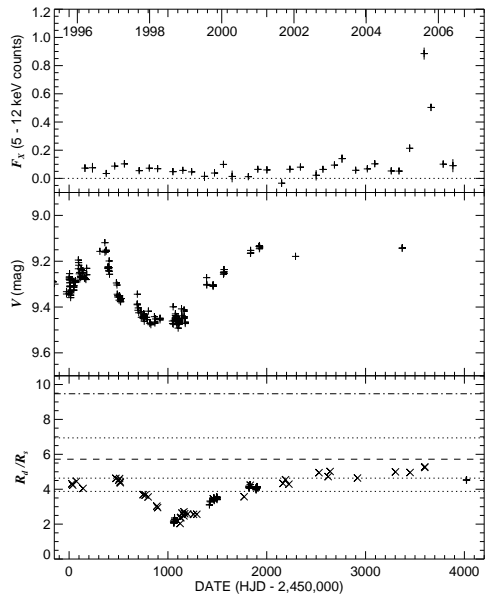}
\caption{The time evolution of the X-ray flux ({\it top}), 
$V$ (or $J$ proxy) magnitude ({\it middle}), and 
disk radius ({\it bottom}) of HDE~245770.  The disk radii are derived 
from the H$\alpha$ equivalent widths and estimated $V$ magnitudes
({\it plus signs} for data from Table~1 and {\it crosses} 
for values from earlier work).
The dotted lines indicate disk radii associated with the 
historic mean levels of emission identified by \citet{lyu00}.
The dashed line indicates the disk truncation radius for 
$n=5$, and the dot-dashed line at the top marks a 
radius equal to the periastron separation.}
\label{fig5}
\end{figure}

\placefigure{fig5}     


\citet{coe06} also present a time evolution diagram of disk radius 
(see their Fig.~7) that they derive from the velocity separation 
$\triangle V$ of the violet and red peaks of the H$\alpha$ profile 
\citep{hua72}.  Their method leads to somewhat smaller radii than our HWHM radii 
($\approx 83\%$ as large using $\triangle V$ measurements derived from Table~1) 
because the radius derived from $\triangle V$ is based upon a velocity sampling 
weighted by the brighter parts of the disk while the HWHM 
radius corresponds to an outer boundary between the bright and faint parts of 
the disk.  The two depictions of disk evolution are in general agreement 
except for estimates from 2005 August where Coe et al.\ determine a 
disk radius about twice as large as indicated in our Figure~5.
We think this discrepancy is due to the fact Coe et al.\ 
relied on an extrapolation of a linear relationship between 
$W_\lambda[{\rm H}\alpha]$ and $\triangle V$ to estimate radius
for the 2005 August data (for which no direct $\triangle V$ measurements
were available).  If one adopts a standard power law
relation between these variables \citep{zam01} rather than a linear 
fit (which tends to underestimate $\triangle V$ for strong emission), 
then the extrapolation leads to radii similar to those shown in Figure~5
for 2005 August.  

We also show in the lower panel of Figure~5 dotted lines that
indicate disk HWHM radii derived from mean levels of H$\alpha$ strength 
in the past that were highlighted in the review by \citet{lyu00}
(and indicated in their Fig.~3).
The lower two levels correspond to the two groupings of 
H$\alpha$ strength observed between 1987 and 1998 while the 
upper level shows the much larger disk HWHM radius associated with 
data from 1975 through 1981 when the H$\alpha$ emission was 
exceptionally strong ($W_\lambda \approx -26$ \AA , \citealt{del84};
$V\approx 8.9$, \citealt{lyu00}).
Note that the deceleration in disk growth observed in 
2001 occurred at the time when the disk HWHM radius had reached 
the lower cluster level.  

\citet{hai04} and \citet{coe06} argue that the tendency for the disk 
emission fluxes to cluster at specified levels is related to the 
presence of resonances between the disk gas and neutron star 
orbital periods that tend to truncate the disk at specific disk 
radii \citep{oka01}.   These truncation radii are given by 
\begin{equation}
{R_n \over R_s} 
 = \left({{GM_s}\over{4\pi^2}}\right)^{1/3} {1\over R_s} \left({P_{\rm orbit} \over n}\right)^{2/3}
 = {{r (M_s/M_\odot )^{1/3}} \over {n^{2/3} R_s/R_\odot}} 
\end{equation}
where $n$ is the integer number of gas rotation periods per neutron star
orbit and $r$ is a constant equal to 97 if the resonance occurs with 
the orbital period or 92 if the resonance is with the shorter beat 
period ($103$~d; \citealt*{lar01}) presumably caused by retrograde precession 
of the disk \citep{hai04}.  \citet{oka01} predict that the important
truncation radii for HDE~245770 will be those associated with the 
$n=4$ and 5 resonances, and using their adopted parameters 
(Table~2) the resonance radii will occur at 
$R_d / R_s = 6.6$ and 5.7 for $n=4$ and 5, respectively
(for $r=92$).  The $n=5$ truncation radius is indicated as a dashed line in 
the lower panel of Figure~5, and we see that the disk HWHM radius
hovered slightly below this limit between 2001 and 2006.  
This suggests that the rapid disk growth phase ended  
when the outer parts of the disk reached the $n=4$ or 5 resonance
truncation radius. 

The top panel of Figure~5 shows the mean X-ray fluxes\footnote{http://xte.mit.edu}  
binned in time slots equal to one orbital period, 
and error bars indicate the standard deviation of the mean within 
the time bin.  The slow expansion of the 
disk apparently led to the large X-ray outburst observed near 
JD~2,453,526 \citep{coe06} and a second smaller outburst near JD~2,453,616. 
These outbursts occurred when the disk had reached its largest HWHM radius
over the duration of the {\it RXTE}/ASM mission.  
The binary semimajor axis is approximately $17.9 R_s$ (Table~2)
and the periastron separation is $9.5 R_s$  
(shown as the {\it dot-dashed line} in lower panel of Fig.~5).  
The mean Roche radius of the Be star will 
be approximately $5.7 R_s$ at periastron, about the size of disk HWHM radius
at the time of the outbursts.  This suggests that the disk had grown  
to a size sufficient to permit active mass transfer  
at periastron around the epoch when the outbursts were observed. 
Furthermore, the huge emission strengths observed during 1976 -- 1981 
indicate that the disk may have attained a HWHM radius of $R_d / R_s = 9$ then 
({\it top dotted line} in the lower panel of Fig.~5), exceeding the Roche limit
at periastron. 

We suspect that the mass transfer process is probably aided 
by the strong tidal forces that exist near periastron.  
\citet{oka02} present hydrodynamical simulations for 
BeXRBs that show how the periastron tides excite a two-arm spiral 
structure in the Be star disk, and the arm closest to the neutron 
star can lift material far beyond the nominal disk boundary 
to provide a source of gas accretion at phases beyond periastron. 
We show in Figure~6 the X-ray light curves binned in orbital 
phase according to the ephemeris of \citet{fin94} 
for the three energy bands observed by {\it RXTE}/ASM.  
The time sample is restricted to quiescent (non-outburst) dates.  
We see that the high energy flux attains a maximum near 
orbital phase 0.3, i.e., well past periastron at phase 0.0. 
A similar phase delay in the emission strength of H$\beta$ 
was detected by \citet{mot91}.  These observations suggest 
that the disk does become more extended for a period 
following periastron, leading to a delayed peak in mass transfer 
and X-ray emission.  

\begin{figure}
\plotone{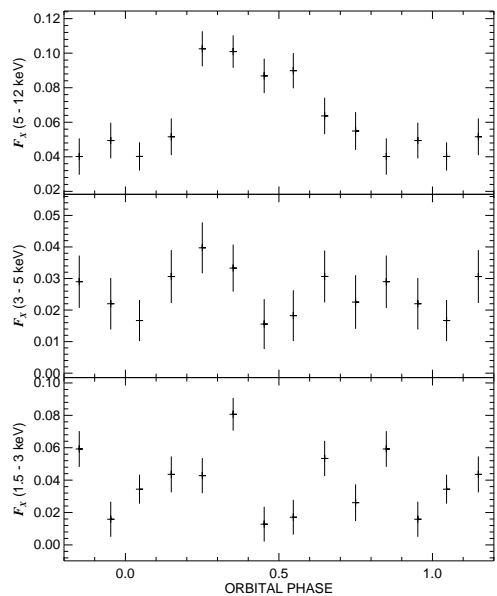} 
\caption{X-ray light curves of HDE~245770 from {\it RXTE}/ASM 
observations during quiescent dates formed by binning with 
the orbital ephemeris from \citet{fin94}.  The high energy band 
({\it top panel}) appears to attain a maximum at a phase 0.3 
past periastron.}
\label{fig6}
\end{figure}

\placefigure{fig6}     

\subsection{X Persei} 

The H$\alpha$ emission equivalent widths of X~Per observed in 
2000 and 2004 and especially in 2006 are the largest measured over 
the last few decades \citep{roc93,pib00,cla01}, and they indicate that 
the disk has grown significantly in size.  Once again 
we need to rescale the equivalent widths to a constant 
stellar continuum flux level by considering the $(1 + \epsilon)$ factor. 
We estimated this flux excess using the $V$-band light curve 
and the relation  
\begin{equation}
(1+\epsilon) ~W_\lambda = 10^{-0.4(V-6.78)} ~W_\lambda
\end{equation}
where the base level magnitude of the star alone, $V=6.78$, 
is that observed during the last disk-free phase \citep{lyu97}. 
The $V$-band light curve comes from observations by the 
members of the American Association of Variable Star Observers 
(AAVSO) that we transformed to standard Johnson $V$ magnitude 
using contemporaneous photometry from \citet{zam95}, \citet{eng97}, 
and photoelectric measurements from the AAVSO \citep{per01}. 
The transformed AAVSO measurements were binned into 25~d means 
to increase the S/N ratio.  

Our results are plotted as function of time in Figure~7. 
The lower panel shows the time evolution of the disk HWHM radius 
derived from our H$\alpha$ equivalent width measurements 
and other published values \citep{cla01,liu01,zam01}. 
The middle panel shows the time binned $V$-band light curve
from the AAVSO observations.  We see that X~Per reached 
a maximum of $V\approx 6.2$ in 2000 and then 
rose again to $V\approx 6.1$ where it has remained to the present.  
The circumstellar disk of X~Per is now brighter than it has been 
for the past few decades \citep{roc93}.   
The top panel shows the binned X-ray fluxes
observed with {\it RXTE}/ASM.  We see that the  
first brightening episode corresponded to the onset of a slow 
increase in X-ray flux that eventually soared to a record high 
over the ASM observation period shortly after the second 
visual brightening.  The X-ray flux continues to remain high
during the current optically bright state. 

\begin{figure}
\plotone{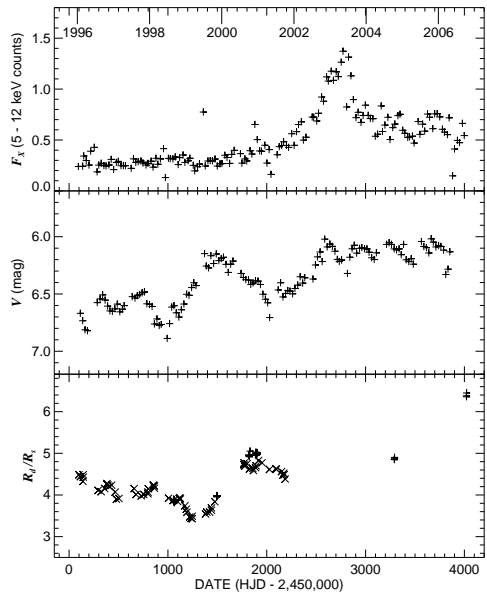}
\caption{The time variations in 
X-ray flux 
({\it top panel:} {\it RXTE}/ASM counts binned in 25~d increments),
$V$ magnitude 
({\it middle panel:} AAVSO observations binned in 25~d increments),
and the ratio of disk to stellar radius 
({\it lower panel:} based upon the rescaled H$\alpha$ equivalent width). 
The plus symbols in the lower plot indicate radii derived 
from our equivalent width measurements, while the crosses
are other published measurements.}
\label{fig7}
\end{figure}

\placefigure{fig7}     

The apparent increase in X-ray flux (and implied gas accretion rate) 
that accompanied the disk expansion confirms 
that the X-ray source is powered by gas from the 
Be star disk.   However, the largest disk HWHM radius shown in Figure~7 
is still much smaller than the separation between the Be star 
and neutron star.  For example, the mean Roche radius
at periastron is $34 R_S$ (Table~2), which is a factor of five larger 
than the maximum disk HWHM radius ($R_d / R_s = 6.4$ in 
Fig.~7).  The maximum predicted disk size due to the action 
of the tidally-driven eccentric instability occurs at the $3:1$ 
resonance radius between the disk gas and neutron star orbital
periods \citep{oka01,cla01}, and this radius occurs at 
$R_d / R_s = 31$ (Table~2).  Once again the observed maximum HWHM radius 
is smaller than the predicted truncation limit. 

However, tidal forces at periastron may excite a two-armed spiral in the 
disk, and the spiral arm facing the companion can lift disk 
gas out to close to the vicinity of the neutron star \citep{oka02}.  
If so, mass transfer will occur predominantly after periastron 
as the gas in the arm extension is accreted by the neutron star. 
We show in Figure~8 the X-ray light curves for X~Per from fluxes recorded in 
the recent active state (HJD $>$ 2,452,200) that are binned according 
to the orbital ephemeris of \citet{del01}.  We see that there is 
an orbital modulation (particularly in the high energy band) that was not 
obvious in the earlier, low state data \citep{wen06} and that the 
X-ray maximum occurs about one quarter of a period after periastron. 
This phase of maximum is consistent with post-periastron accretion 
from extended disk gas in a spiral arm. 

\begin{figure}
\plotone{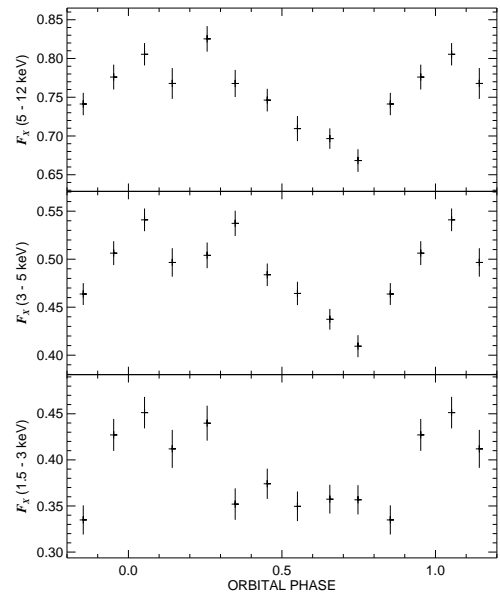} 
\caption{X-ray light curves of X~Per from {\it RXTE}/ASM 
observations during active dates formed by binning with 
the orbital ephemeris from \citet{del01}.  The high energy band 
({\it top panel}) appears to attain a maximum at a phase 
$\approx 0.25$ past periastron.  The vertical segments indicate 
the $\pm 1 \sigma$ standard deviation of the mean within each bin.}
\label{fig8}
\end{figure}

\placefigure{fig8}     


\section{Discussion}                                

The H$\alpha$ observations presented here and elsewhere 
document the remarkable and continuing changes that occur in the 
mass loss from the Be stars in these two BeXRB systems.  
The gas that enters their circumstellar disks becomes a 
reservoir of fuel for accretion by the neutron star companion, 
and the X-ray activity that accompanies the mass transfer 
increases dramatically as the disk radius increases in size.
The disk growth is expected to be limited by gravitational 
interactions with the neutron star \citep{oka01,oka02}. 
In low eccentricity binaries like X~Per, the limiting radius
occurs at the 3:1 resonance radius by the tidally-driven 
eccentric instability, while in high eccentricity systems 
like HDE~245770, the neutron star will approach closer to 
the Be star at periastron and higher integer resonances 
occurring at smaller radii will act to truncate the disk. 

Our observations appear to confirm these expectations in 
the case of HDE~245770.  The largest disk HWHM radius we find 
from the H$\alpha$ equivalent width is comparable to both 
the $n=5$ resonance radius and the mean Roche radius at 
the time of periastron.  The historical maxima of H$\alpha$ 
strength may imply that the disk HWHM radius can occasionally grow to even 
larger dimensions.  We caution, however, that it may 
be possible for Be disks of extreme density to attain very 
large H$\alpha$ emission fluxes without radial growth. 
For example, pressure broadening in high density disks 
can lead to an increased flux in the wings of H$\alpha$ 
and thus a larger emission equivalent width.   
Whether such high density and high pressure disks could 
remain confined by gravitational forces is an open question. 

The case of X~Per, on the other hand, shows that the mass 
transfer and X-ray accretion flux can increase even when the 
disk HWHM radius attains a size well below the critical value. 
We found that the recent X-ray flux increase occurred 
when the disk HWHM radius grew to about $R_d/R_s=5$, much less 
than the resonant truncation radius of $R_3/R_s=31$.  There are two 
probable explanations for this difference.    
First, tidal forces at periastron promote the development 
of a two-arm spiral structure in the disk \citep{oka02} 
and gas concentrated in the arm closest to the companion
can be pushed outwards and accreted by the neutron star. 
Hydrodynamical models suggest that the gas accretion rate from 
the spiral arm will peak sometime after periastron, and 
we find that indeed the X-ray flux maximum lags periastron by 
$\approx 25\%$ of the orbital period in both the BeXRBs 
examined here and in the microquasar LS~I~+61~303 \citep{gr07b}.
Secondly, the disk probably extends beyond the HWHM radius of 
the H$\alpha$ emission that we use to define disk radius \citep{grg06}. 
The disk density is described in our model in terms of a power law 
with radial distance from the star \citep{hum00}, and the lower 
gas density at larger radius may be entirely adequate to power 
the accretion-driven X-ray flux. 

The observational record for both of these BeXRBs documents
the growth of the disk from inside out.  A comparison of the 
$V$-band light curves and H$\alpha$ radii in both Figures~5 
and 7 shows that the time evolution of the H$\alpha$ emission 
lags behind that of the disk continuum flux.  \citet{cla01} also 
noted this delay between the line and continuum fluxes, and  
they argue that the time lag is due to differences in spatial origin.   
The continuum flux excess is probably formed in the inner 
part of the disk while the H$\alpha$ emission is more 
optically thick and forms over a wider range of disk radius. 
Thus, an outwards propagating density enhancement would 
peak first in continuum light and later in the H$\alpha$ flux. 
The fate of this outflow is unclear.  Models by \citet{oka02} 
suggest that some of the mass falls back onto the Be star, 
some is accreted by the neutron star, and the remainder 
escapes from the system (presumably into a circumbinary region 
centered on the orbital plane).   A search for such 
circumstellar structures might yield evidence of the ejected 
gas (for example, as a compact H~II region surrounding X~Per;
\citealt{rey05}).


\acknowledgments

We thank Daryl Willmarth and the staff of KPNO for their assistance 
in making these observations possible.  We acknowledge with thanks 
the variable star observations from the AAVSO International Database 
contributed by observers worldwide and used in this research.
The X-ray results were provided by the ASM/RXTE teams at 
MIT and at the RXTE SOF and GOF at NASA's GSFC. 
The {\it IUE} data presented in this paper were obtained from 
the Multimission Archive at the Space Telescope Science 
Institute (MAST). STScI is operated by the Association of 
Universities for Research in Astronomy, Inc., under NASA 
contract NAS5-26555. Support for MAST for non-HST data is 
provided by the NASA Office of Space Science via grant 
NAG5-7584 and by other grants and contracts.
This work was supported by the National Science 
Foundation under grants AST-0205297, AST-0506573, and AST-0606861.
Institutional support has been provided from the GSU College
of Arts and Sciences and from the Research Program Enhancement
fund of the Board of Regents of the University System of Georgia,
administered through the GSU Office of the Vice President
for Research.  



\clearpage
\appendix 

\section{Mass Ratio Limit from {\it IUE} Radial Velocities of X Per}

Several investigators have attempted to measure the orbital motion 
of X~Per, but their results have generally been inconclusive
because the lines are broad, shallow, and often marred by emission
and because the semiamplitude is probably small \citep{hut77,rey92,sti92}.  
The ideal spectral range to search for orbital motion is in 
the ultraviolet because the disk flux contribution is relatively small in 
the UV \citep{tel98}.  \citet{sti92} used the spectra available
at that time in the archive of the {\it IUE} to measure radial 
velocities, and he found that the velocity excursions were too 
small to measure orbital motion.  Here we repeat the analysis of 
the now larger set of {\it IUE} spectra using the complementary 
orbital elements from the pulsar orbit \citep{del01}.  

We obtained 43 high dispersion, short wavelength prime camera 
spectra of X~Per from the {\it IUE} database 
maintained at the Multimission Archive at Space 
Telescope\footnote{http://archive.stsci.edu/iue/}. 
These spectra were transformed to a uniform $\log \lambda$,  
heliocentric wavelength grid, normalized to a 
pseudo-continuum, and excised of the main interstellar
lines, and radial velocities were measured by cross-correlation 
\citep*{pen99}.  A mean formed from the average of all 
the spectra was used as a spectral template, and its 
absolute velocity was determined to be $+0.8$ km~s$^{-1}$ 
by cross-correlation with a similar spectrum of 
HD~34078 \citep{gie86}.  We used the apparent shifts 
between each cross-correlation function and the ensemble 
mean of the functions to estimate the relative velocity 
shift for each spectrum, and then the final absolute velocities
were set by adding the template velocity above.  The 
results are listed in Table~3 (given in full in the 
electronic version) that gives the 
heliocentric Julian date of mid-exposure, the orbital 
phase from the ephemeris of \citet{del01}, and the 
measured radial velocity. 

\placetable{tab3}      

We then fit the {\it IUE} velocities using the 
non-linear, least-squares, orbital elements program 
of \citet{mor74}.   This was a constrained solution in 
which all the elements were set from the pulsar results
\citep{del01} (with the longitude of periastron changed 
by $180^\circ$) with the exception of the systemic velocity 
$\gamma$ and the semiamplitude $K$.   The formal solution 
for these parameters is $\gamma = 1.0 \pm 0.9$ km~s$^{-1}$
and $K = 2.3 \pm 1.4$ km~s$^{-1}$, and the observations 
and calculated radial velocity curve are illustrated in 
Figure~9.   Clearly the amplitude of motion is small 
enough that this result is of marginal significance, 
but the velocities do place a useful upper limit on the 
semiamplitude.  The dotted line in Figure~9 shows the 
fit with a semiamplitude of $K + 2\sigma = 5.0$ km~s$^{-1}$, 
and this appears to be a reasonable upper limit 
consistent with the spread in the {\it IUE} velocities. 
 
\begin{figure}
\begin{center}
{\includegraphics[angle=90,height=12cm]{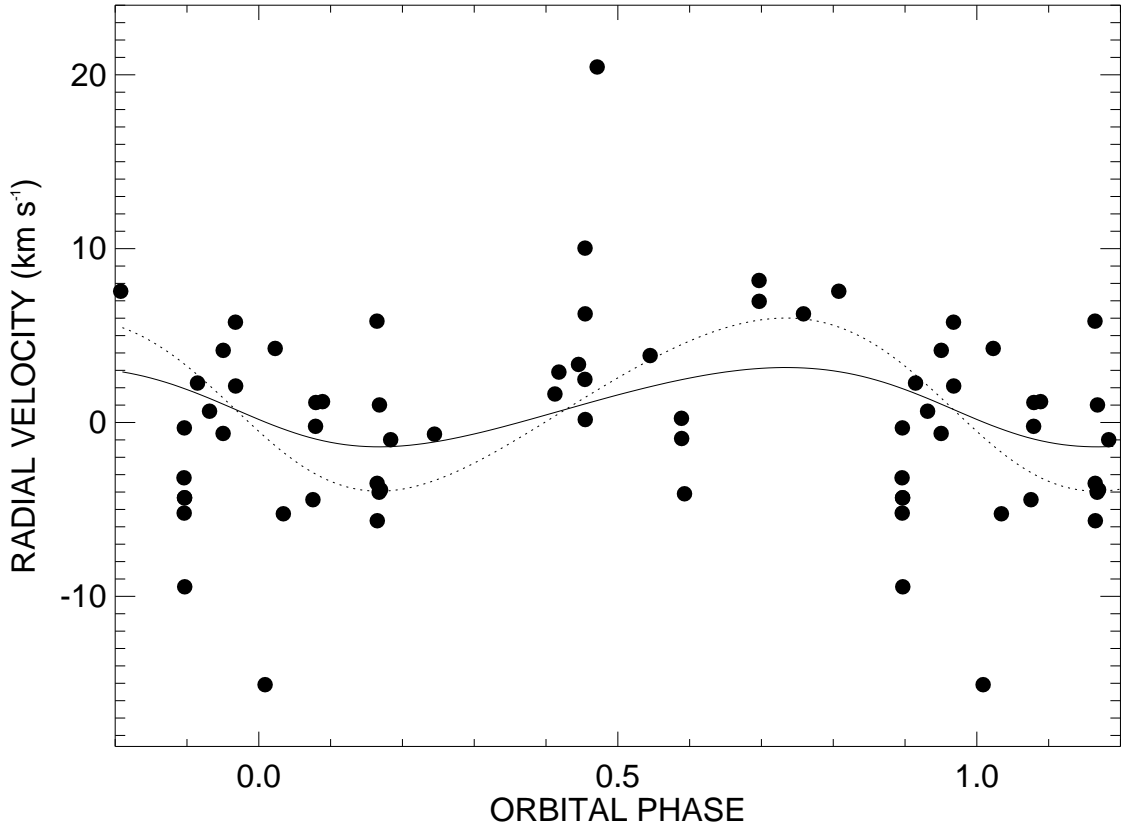}}
\end{center}
\caption{Radial velocities of X~Per from {\it IUE} spectroscopy 
plotted against the orbital ephemeris of the X-ray pulsar. 
The solid line shows the nominal best fit of the Be star 
systemic velocity and semiamplitude (with all other parameters 
set by the pulsar orbit) while the dotted line shows a fit 
with the semiamplitude fixed at the $2\sigma$ upper limit.}
\label{fig9}
\end{figure}

\placefigure{fig9}     

The pulsar orbital elements \citep{del01} give a neutron star 
orbital semiamplitude of $K=39.8 \pm 0.4$ km~s$^{-1}$, 
so the ratio of the upper limit of the Be star semiamplitude 
to that for the pulsar yields an upper limit on the 
mass ratio, $M_2/M_1 < 0.13$.  We have plotted this 
constraint in the mass plane diagram in Figure~10 together 
with lines of constant orbital inclination derived from the 
pulsar mass function.   A lower limit on the inclination 
of $i> 23^\circ$ results from assuming that the star rotates 
slower than the critical rate \citep{cla01}.  \citet{har88}
presents a summary of mass and radii data from eclipsing 
binary stars, and the mass range for a B0~V star is probably 
between $\approx 11 M_\odot$ and $17 M_\odot$.  The shaded region 
in Figure~10 shows this probable Be star range together with 
the observed range in neutron star mass \citep*{van95}. 
This region in the mass plane corresponds to an inclination range of 
$i=28^\circ$ to $35^\circ$. 

\begin{figure}
\plotone{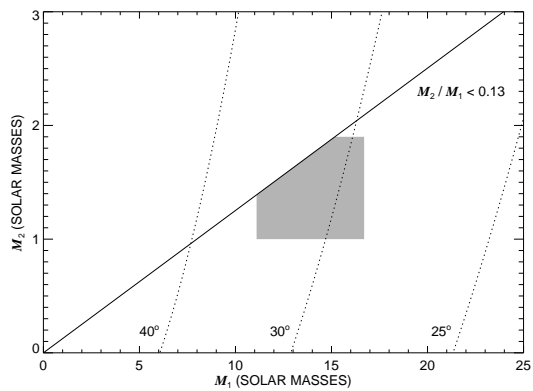}
\caption{A mass plane diagram for X~Per showing constraints on the 
Be star mass $M_1$ and the neutron star mass $M_2$. 
The solid line shows the upper limit on mass ratio determined 
from the {\it IUE} radial velocities, and the dotted lines 
show the mass relations for three values of orbital inclination.
The shaded region shows the most probable range in masses 
for both components.}
\label{fig10}
\end{figure}

\placefigure{fig10}     




\clearpage


\begin{deluxetable}{lcccccccccc}
\rotate
\tabletypesize{\scriptsize}
\tablewidth{0pt}
\tablecaption{H$\alpha$ Measurements\label{tab1}}
\tablehead{
\colhead{Star} &
\colhead{Date} &
\colhead{Orbital} &
\colhead{$W_\lambda$}  &
\colhead{$V_r (W)$} &
\colhead{$V_r (V)$} &
\colhead{$V_r (R)$} &
\colhead{$W_\lambda (V)~/$} &
\colhead{} &
\colhead{FWHM($V$)} &
\colhead{FWHM($R$)}
\\
\colhead{Name} &
\colhead{(HJD$-$2,400,000)} &
\colhead{Phase} &
\colhead{(\AA )} &
\colhead{(km s$^{-1}$)} &
\colhead{(km s$^{-1}$)} &
\colhead{(km s$^{-1}$)} &
\colhead{$W_\lambda (R)$} &
\colhead{$V/R$} &
\colhead{(\AA )} &
\colhead{(\AA )} 
}
\startdata
HDE 245770\dotfill & 51055.969 &  0.103 & \phn \phs$1.76$ &
   \phs$18.4$ & \nodata &     \nodata & \nodata & \nodata & \nodata & \nodata \\
HDE 245770\dotfill & 51056.989 &  0.112 & \phn \phs$1.96$ &
   \phs$42.4$ & \nodata &     \nodata & \nodata & \nodata & \nodata & \nodata \\
HDE 245770\dotfill & 51057.960 &  0.121 & \phn \phs$1.41$ &
   \phs$32.4$ & \nodata &     \nodata & \nodata & \nodata & \nodata & \nodata \\
HDE 245770\dotfill & 51058.950 &  0.130 & \phn \phs$2.16$ &
   \phs$30.5$ & \nodata &     \nodata & \nodata & \nodata & \nodata & \nodata \\
HDE 245770\dotfill & 51061.954 &  0.157 & \phn \phs$1.96$ &
   \phs$38.0$ & \nodata &     \nodata & \nodata & \nodata & \nodata & \nodata \\
HDE 245770\dotfill & 51061.993 &  0.158 & \phn \phs$2.07$ &
   \phs$31.7$ & \nodata &     \nodata & \nodata & \nodata & \nodata & \nodata \\
HDE 245770\dotfill & 51063.962 &  0.176 & \phn \phs$1.62$ &
   \phs$21.1$ & \nodata &     \nodata & \nodata & \nodata & \nodata & \nodata \\
HDE 245770\dotfill & 51065.981 &  0.194 & \phn \phs$1.05$ &
   \phs$16.5$ & \nodata &     \nodata & \nodata & \nodata & \nodata & \nodata \\
HDE 245770\dotfill & 51066.902 &  0.202 & \phn \phs$0.52$ &
   \phs$18.0$ & \nodata &     \nodata & \nodata & \nodata & \nodata & \nodata \\
HDE 245770\dotfill & 51066.976 &  0.203 & \phn \phs$0.88$ &
   \phs$20.3$ & \nodata &     \nodata & \nodata & \nodata & \nodata & \nodata \\
HDE 245770\dotfill & 51419.972 &  0.403 & \phn  $-$3.11 &
   \phs$22.5$ & $-147.7$ &     165.5 & 0.80 & 0.96 & 4.41 & 5.34 \\
HDE 245770\dotfill & 51419.995 &  0.403 & \phn  $-$3.20 &
   \phs$23.1$ & $-144.9$ &     165.6 & 0.80 & 0.95 & 4.68 & 5.52 \\
HDE 245770\dotfill & 51425.987 &  0.458 & \phn  $-$4.31 &
\phn\phs$8.6$ & $-158.3$ &     161.6 & 0.95 & 0.97 & 5.12 & 5.26 \\
HDE 245770\dotfill & 51428.950 &  0.485 & \phn  $-$4.06 &
   \phs$12.1$ & $-151.8$ &     165.0 & 0.96 & 0.98 & 4.97 & 5.10 \\
HDE 245770\dotfill & 51464.958 &  0.811 & \phn  $-$5.32 &
\phn\phs$4.5$ & $-157.6$ &     156.7 & 0.98 & 1.06 & 4.84 & 5.24 \\
HDE 245770\dotfill & 51464.980 &  0.811 & \phn  $-$5.14 &
\phn\phs$4.7$ & $-155.8$ &     158.8 & 1.00 & 1.05 & 4.89 & 5.14 \\
HDE 245770\dotfill & 51465.936 &  0.820 & \phn  $-$5.17 &
\phn\phs$3.1$ & $-156.5$ &     154.2 & 0.99 & 1.05 & 4.75 & 5.04 \\
HDE 245770\dotfill & 51465.957 &  0.820 & \phn  $-$5.04 &
\phn\phs$2.1$ & $-156.6$ &     153.9 & 0.99 & 1.03 & 4.81 & 5.01 \\
HDE 245770\dotfill & 51466.906 &  0.829 & \phn  $-$4.80 &
\phn   $-3.6$ & $-157.5$ &     149.1 & 1.02 & 1.04 & 4.72 & 4.78 \\
HDE 245770\dotfill & 51466.928 &  0.829 & \phn  $-$4.79 &
\phn   $-5.9$ & $-158.5$ &     149.3 & 1.06 & 1.03 & 4.81 & 4.69 \\
HDE 245770\dotfill & 51467.988 &  0.839 & \phn  $-$4.41 &
\phn   $-3.6$ & $-153.5$ &     155.5 & 1.06 & 1.05 & 4.87 & 4.85 \\
HDE 245770\dotfill & 51468.952 &  0.847 & \phn  $-$4.54 &
\phn   $-4.5$ & $-152.6$ &     159.5 & 1.22 & 1.06 & 5.28 & 4.61 \\
HDE 245770\dotfill & 51468.974 &  0.847 & \phn  $-$4.73 &
\phn   $-1.5$ & $-152.9$ &     155.8 & 1.12 & 1.06 & 5.03 & 4.75 \\
HDE 245770\dotfill & 51469.916 &  0.856 & \phn  $-$4.87 &
\phn   $-8.1$ & $-155.9$ &     152.1 & 1.12 & 1.06 & 5.05 & 4.76 \\
HDE 245770\dotfill & 51491.904 &  0.055 & \phn  $-$5.01 &
\phn\phs$6.0$ & $-148.6$ &     154.3 & 1.01 & 0.97 & 4.96 & 4.78 \\
HDE 245770\dotfill & 51492.833 &  0.064 & \phn  $-$5.20 &
   \phs$10.0$ & $-148.0$ &     153.6 & 0.93 & 1.00 & 4.74 & 5.09 \\
HDE 245770\dotfill & 51493.827 &  0.073 & \phn  $-$5.25 &
   \phs$12.3$ & $-150.5$ &     156.9 & 0.92 & 0.98 & 4.82 & 5.10 \\
HDE 245770\dotfill & 51494.869 &  0.082 & \phn  $-$5.57 &
   \phs$12.8$ & $-147.5$ &     163.4 & 0.95 & 1.01 & 4.76 & 5.08 \\
HDE 245770\dotfill & 51494.890 &  0.082 & \phn  $-$5.36 &
   \phs$10.9$ & $-148.5$ &     156.6 & 0.96 & 1.01 & 4.74 & 4.98 \\
HDE 245770\dotfill & 51495.944 &  0.092 & \phn  $-$5.28 &
\phn\phs$6.7$ & $-148.6$ &     157.8 & 1.01 & 0.99 & 4.92 & 4.84 \\
HDE 245770\dotfill & 51496.919 &  0.101 & \phn  $-$5.20 &
   \phs$12.6$ & $-147.7$ &     161.0 & 0.94 & 1.00 & 4.69 & 4.98 \\
HDE 245770\dotfill & 51497.852 &  0.109 & \phn  $-$4.66 &
   \phs$17.8$ & $-144.8$ &     161.8 & 0.92 & 0.96 & 4.76 & 4.98 \\
HDE 245770\dotfill & 51817.905 &  0.011 & \phn  $-$8.03 &
   \phs$12.4$ & $-126.1$ &     133.5 & 0.92 & 0.96 & 4.77 & 5.00 \\
HDE 245770\dotfill & 51818.908 &  0.020 & \phn  $-$8.36 &
\phn\phs$6.2$ & $-120.7$ &     136.8 & 1.04 & 0.98 & 5.11 & 4.78 \\
HDE 245770\dotfill & 51819.899 &  0.029 & \phn  $-$8.00 &
\phn\phs$8.7$ & $-120.8$ &     136.4 & 1.04 & 1.00 & 4.96 & 4.77 \\
HDE 245770\dotfill & 51821.884 &  0.047 & \phn  $-$7.66 &
\phn\phs$6.6$ & $-121.6$ &     132.3 & 0.99 & 0.98 & 4.83 & 4.77 \\
HDE 245770\dotfill & 51822.888 &  0.056 & \phn  $-$7.83 &
\phn\phs$4.7$ & $-123.3$ &     133.3 & 0.99 & 0.98 & 4.89 & 4.84 \\
HDE 245770\dotfill & 51823.830 &  0.065 & \phn  $-$7.86 &
   \phs$11.9$ & $-119.7$ &     134.9 & 0.98 & 1.00 & 4.84 & 4.90 \\
HDE 245770\dotfill & 51823.934 &  0.066 & \phn  $-$8.22 &
\phn\phs$5.4$ & $-122.1$ &     134.8 & 1.01 & 1.01 & 4.93 & 4.92 \\
HDE 245770\dotfill & 51824.855 &  0.074 & \phn  $-$8.19 &
   \phs$13.4$ & $-121.5$ &     134.8 & 0.94 & 0.96 & 4.90 & 5.01 \\
HDE 245770\dotfill & 51824.962 &  0.075 & \phn  $-$7.86 &
\phn\phs$8.7$ & $-121.9$ &     131.6 & 0.94 & 0.99 & 4.76 & 5.02 \\
HDE 245770\dotfill & 51830.913 &  0.129 & \phn  $-$8.81 &
   \phs$13.2$ & $-115.1$ &     140.6 & 1.09 & 1.03 & 5.15 & 4.88 \\
HDE 245770\dotfill & 51890.840 &  0.672 & \phn  $-$7.21 &
\phn\phs$6.6$ & $-119.4$ &     130.1 & 1.03 & 1.01 & 4.84 & 4.76 \\
HDE 245770\dotfill & 51890.861 &  0.672 & \phn  $-$7.03 &
\phn\phs$7.3$ & $-120.6$ &     129.6 & 1.00 & 0.99 & 4.79 & 4.72 \\
HDE 245770\dotfill & 51892.806 &  0.690 & \phn  $-$6.97 &
\phn\phs$6.1$ & $-119.8$ &     127.1 & 0.98 & 0.99 & 4.79 & 4.83 \\
HDE 245770\dotfill & 51892.827 &  0.690 & \phn  $-$6.99 &
\phn\phs$5.4$ & $-117.8$ &     130.3 & 1.01 & 0.99 & 4.88 & 4.76 \\
HDE 245770\dotfill & 51893.812 &  0.699 & \phn  $-$7.60 &
\phn\phs$6.5$ & $-117.2$ &     131.6 & 1.04 & 0.99 & 4.96 & 4.71 \\
HDE 245770\dotfill & 51893.833 &  0.699 & \phn  $-$7.55 &
\phn\phs$7.2$ & $-114.8$ &     134.4 & 1.10 & 0.99 & 5.03 & 4.56 \\
HDE 245770\dotfill & 51894.799 &  0.708 & \phn  $-$7.61 &
\phn\phs$5.7$ & $-117.9$ &     132.9 & 1.08 & 0.99 & 5.08 & 4.67 \\
HDE 245770\dotfill & 51894.823 &  0.708 & \phn  $-$7.65 &
\phn\phs$5.4$ & $-118.5$ &     132.5 & 1.05 & 0.99 & 4.98 & 4.70 \\
HDE 245770\dotfill & 51895.834 &  0.717 & \phn  $-$7.63 &
\phn\phs$7.5$ & $-118.2$ &     131.4 & 1.04 & 1.00 & 5.00 & 4.80 \\
HDE 245770\dotfill & 51895.913 &  0.718 & \phn  $-$7.53 &
\phn\phs$7.4$ & $-117.5$ &     131.8 & 1.04 & 1.00 & 4.95 & 4.77 \\
HDE 245770\dotfill & 51896.834 &  0.727 & \phn  $-$7.37 &
\phn\phs$7.5$ & $-118.7$ &     130.3 & 1.00 & 0.99 & 4.84 & 4.79 \\
HDE 245770\dotfill & 51896.855 &  0.727 & \phn  $-$7.46 &
\phn\phs$5.7$ & $-117.9$ &     132.1 & 1.05 & 1.00 & 5.00 & 4.75 \\
HDE 245770\dotfill & 51897.832 &  0.736 & \phn  $-$7.47 &
\phn\phs$9.2$ & $-119.2$ &     128.6 & 0.98 & 1.00 & 4.81 & 4.94 \\
HDE 245770\dotfill & 51897.853 &  0.736 & \phn  $-$7.57 &
\phn\phs$8.9$ & $-118.9$ &     129.9 & 1.00 & 1.00 & 4.85 & 4.86 \\
HDE 245770\dotfill & 51898.841 &  0.745 & \phn  $-$7.77 &
\phn\phs$8.0$ & $-118.8$ &     132.2 & 1.03 & 1.00 & 4.89 & 4.77 \\
HDE 245770\dotfill & 51898.862 &  0.745 & \phn  $-$7.78 &
\phn\phs$8.2$ & $-119.5$ &     132.0 & 1.01 & 1.01 & 4.88 & 4.85 \\
HDE 245770\dotfill & 51899.838 &  0.754 & \phn  $-$7.90 &
\phn\phs$8.3$ & $-118.8$ &     132.8 & 1.02 & 0.99 & 4.96 & 4.81 \\
HDE 245770\dotfill & 51899.859 &  0.754 & \phn  $-$7.86 &
\phn\phs$8.5$ & $-118.8$ &     134.2 & 1.03 & 0.99 & 5.00 & 4.80 \\
HDE 245770\dotfill & 51900.831 &  0.763 & \phn  $-$7.95 &
\phn\phs$6.1$ & $-117.8$ &     136.2 & 1.11 & 1.00 & 5.18 & 4.69 \\
HDE 245770\dotfill & 51900.852 &  0.763 & \phn  $-$8.00 &
\phn\phs$6.3$ & $-120.2$ &     135.0 & 1.07 & 1.00 & 5.14 & 4.83 \\
HDE 245770\dotfill & 51901.817 &  0.772 & \phn  $-$7.87 &
\phn\phs$8.3$ & $-115.8$ &     136.5 & 1.11 & 1.00 & 5.18 & 4.65 \\
HDE 245770\dotfill & 51901.838 &  0.772 & \phn  $-$7.79 &
\phn\phs$9.4$ & $-117.2$ &     135.0 & 1.06 & 0.99 & 5.09 & 4.77 \\
HDE 245770\dotfill & 54019.981 &  0.975 &      $-$10.08 & 
\phn\phs$0.5$ & $-119.4$ &     103.1 & 0.82 & 0.82 & 4.92 & 4.94 \\
HDE 245770\dotfill & 54020.953 &  0.984 &      $-$10.21 & 
\phn\phs$2.8$ & $-116.3$ &     105.8 & 0.80 & 0.84 & 4.77 & 4.99 \\
HDE 245770\dotfill & 54021.957 &  0.993 &      $-$10.16 & 
\phn\phs$3.1$ & $-113.7$ &     105.9 & 0.84 & 0.83 & 4.88 & 4.81 \\
X Per\dotfill & 51055.896 &  0.811 & \phn $-$9.38 & 
\phn\phs$4.5$ & \phn$ -73.1$ & \phn     93.0 & 0.79 & 0.92 & 2.71 & 3.16 \\
X Per\dotfill & 51056.966 &  0.815 & \phn $-$9.09 & 
\phn\phs$4.6$ & \phn$ -72.3$ & \phn     94.5 & 0.77 & 0.91 & 2.71 & 3.22 \\
X Per\dotfill & 51057.948 &  0.819 & \phn $-$9.64 & 
\phn\phs$4.4$ & \phn$ -71.2$ & \phn     93.5 & 0.82 & 0.92 & 2.74 & 3.09 \\
X Per\dotfill & 51058.938 &  0.823 & \phn $-$9.50 & 
\phn\phs$4.0$ & \phn$ -70.4$ & \phn     95.1 & 0.85 & 0.92 & 2.81 & 3.06 \\
X Per\dotfill & 51061.986 &  0.835 & \phn $-$9.51 & 
\phn\phs$0.6$ & \phn$ -67.6$ & \phn     92.5 & 0.99 & 0.93 & 2.92 & 2.74 \\
X Per\dotfill & 51062.915 &  0.839 & \phn $-$9.49 & 
\phn\phs$0.8$ & \phn$ -69.1$ & \phn     91.1 & 0.93 & 0.93 & 2.84 & 2.85 \\
X Per\dotfill & 51063.954 &  0.843 & \phn $-$9.09 & 
\phn\phs$1.2$ & \phn$ -68.9$ & \phn     92.5 & 0.90 & 0.90 & 2.85 & 2.85 \\
X Per\dotfill & 51065.936 &  0.851 & \phn $-$9.25 & 
\phn\phs$0.7$ & \phn$ -67.2$ & \phn     92.0 & 1.00 & 0.90 & 2.96 & 2.66 \\
X Per\dotfill & 51066.937 &  0.855 & \phn $-$8.99 & 
\phn   $-1.1$ & \phn$ -68.1$ & \phn     90.4 & 0.96 & 0.90 & 2.86 & 2.70 \\
X Per\dotfill & 51491.878 &  0.553 &     $-$10.05 & 
      $-15.1$ &     $-116.4$ & \phn     62.0 & 0.71 & 0.77 & 3.65 & 3.97 \\
X Per\dotfill & 51492.818 &  0.557 &     $-$10.15 & 
      $-11.8$ &     $-113.5$ & \phn     64.2 & 0.70 & 0.77 & 3.58 & 3.94 \\
X Per\dotfill & 51493.813 &  0.561 &     $-$10.24 & 
      $-10.6$ &     $-110.2$ & \phn     67.6 & 0.76 & 0.80 & 3.72 & 3.91 \\
X Per\dotfill & 51494.845 &  0.565 &     $-$10.31 & 
      $-12.5$ &     $-111.4$ & \phn     65.9 & 0.77 & 0.80 & 3.74 & 3.91 \\
X Per\dotfill & 51494.853 &  0.565 &     $-$10.27 & 
      $-12.4$ &     $-112.3$ & \phn     65.5 & 0.75 & 0.80 & 3.68 & 3.92 \\
X Per\dotfill & 51495.873 &  0.569 &     $-$10.30 & 
      $-11.7$ &     $-109.0$ & \phn     67.8 & 0.78 & 0.82 & 3.67 & 3.85 \\
X Per\dotfill & 51496.859 &  0.573 &     $-$10.21 & 
      $-13.5$ &     $-113.6$ & \phn     63.0 & 0.71 & 0.80 & 3.55 & 4.03 \\
X Per\dotfill & 51497.830 &  0.577 &     $-$10.22 & 
      $-12.9$ &     $-115.1$ & \phn     61.7 & 0.68 & 0.79 & 3.52 & 4.09 \\
X Per\dotfill & 51817.899 &  0.855 &     $-$18.63 & 
   \phs$17.5$ & \phn$ -42.7$ &         118.6 & 2.29 & 2.05 & 3.17 & 2.83 \\
X Per\dotfill & 51818.900 &  0.859 &     $-$18.69 & 
   \phs$16.0$ & \phn$ -43.1$ &         118.0 & 2.30 & 2.10 & 3.14 & 2.87 \\
X Per\dotfill & 51819.887 &  0.863 &     $-$18.54 & 
   \phs$15.7$ & \phn$ -43.0$ &         115.8 & 2.34 & 2.09 & 3.13 & 2.79 \\
X Per\dotfill & 51820.918 &  0.867 &     $-$18.59 & 
   \phs$15.7$ & \phn$ -43.5$ &         116.3 & 2.31 & 2.13 & 3.13 & 2.88 \\
X Per\dotfill & 51821.873 &  0.871 &     $-$18.42 & 
   \phs$14.5$ & \phn$ -43.5$ &         116.4 & 2.41 & 2.15 & 3.12 & 2.79 \\
X Per\dotfill & 51822.875 &  0.875 &     $-$18.49 & 
   \phs$14.7$ & \phn$ -42.5$ &         115.8 & 2.40 & 2.12 & 3.13 & 2.76 \\
X Per\dotfill & 51823.823 &  0.879 &     $-$18.26 & 
   \phs$14.5$ & \phn$ -42.3$ &         116.1 & 2.47 & 2.15 & 3.15 & 2.74 \\
X Per\dotfill & 51824.852 &  0.883 &     $-$18.53 & 
   \phs$15.8$ & \phn$ -41.6$ &         115.6 & 2.37 & 2.09 & 3.15 & 2.77 \\
X Per\dotfill & 51824.959 &  0.884 &     $-$18.39 & 
   \phs$16.7$ & \phn$ -40.9$ &         116.3 & 2.37 & 2.08 & 3.15 & 2.76 \\
X Per\dotfill & 51830.954 &  0.908 &     $-$19.59 & 
   \phs$11.9$ & \phn$ -49.3$ &         105.4 & 1.95 & 1.89 & 3.04 & 2.96 \\
X Per\dotfill & 51888.787 &  0.139 &     $-$18.93 & 
\phn\phs$7.8$ & \phn$ -59.0$ &         104.3 & 1.75 & 1.85 & 3.09 & 3.26 \\
X Per\dotfill & 51888.792 &  0.139 &     $-$18.95 & 
\phn\phs$8.9$ & \phn$ -58.5$ &         104.9 & 1.71 & 1.84 & 3.07 & 3.31 \\
X Per\dotfill & 51889.738 &  0.142 &     $-$18.54 & 
\phn\phs$8.3$ & \phn$ -59.1$ & \phn     99.9 & 1.66 & 1.86 & 3.01 & 3.38 \\
X Per\dotfill & 51889.872 &  0.143 &     $-$19.32 & 
\phn\phs$7.5$ & \phn$ -58.4$ &         101.1 & 1.74 & 1.89 & 3.04 & 3.30 \\
X Per\dotfill & 51890.733 &  0.146 &     $-$19.21 & 
\phn\phs$7.7$ & \phn$ -58.4$ &         101.0 & 1.73 & 1.91 & 3.02 & 3.34 \\
X Per\dotfill & 51892.756 &  0.154 &     $-$18.91 & 
\phn\phs$5.4$ & \phn$ -59.6$ &         100.0 & 1.83 & 1.91 & 2.99 & 3.12 \\
X Per\dotfill & 51893.769 &  0.158 &     $-$19.00 & 
\phn\phs$7.6$ & \phn$ -59.2$ &         102.6 & 1.74 & 1.91 & 2.98 & 3.26 \\
X Per\dotfill & 51894.767 &  0.162 &     $-$18.76 & 
\phn\phs$9.1$ & \phn$ -57.7$ &         102.8 & 1.76 & 1.80 & 2.99 & 3.07 \\
X Per\dotfill & 51895.824 &  0.167 &     $-$18.64 & 
\phn\phs$9.0$ & \phn$ -55.2$ &         103.7 & 1.84 & 1.81 & 3.06 & 3.01 \\
X Per\dotfill & 51896.797 &  0.171 &     $-$19.03 & 
\phn\phs$8.7$ & \phn$ -55.0$ &         103.7 & 1.92 & 1.83 & 3.08 & 2.94 \\
X Per\dotfill & 51897.797 &  0.175 &     $-$18.85 & 
\phn\phs$8.2$ & \phn$ -55.5$ &         104.3 & 1.91 & 1.85 & 3.07 & 2.98 \\
X Per\dotfill & 51898.806 &  0.179 &     $-$18.97 & 
\phn\phs$8.4$ & \phn$ -55.0$ &         105.2 & 1.93 & 1.85 & 3.10 & 2.97 \\
X Per\dotfill & 51899.804 &  0.183 &     $-$18.93 & 
\phn\phs$8.7$ & \phn$ -55.3$ &         104.1 & 1.91 & 1.84 & 3.07 & 2.96 \\
X Per\dotfill & 51900.797 &  0.187 &     $-$18.83 & 
\phn\phs$8.9$ & \phn$ -54.0$ &         105.5 & 1.95 & 1.89 & 3.09 & 2.99 \\
X Per\dotfill & 51901.783 &  0.191 &     $-$19.26 & 
\phn\phs$9.0$ & \phn$ -52.4$ &         107.2 & 2.06 & 1.97 & 3.07 & 2.93 \\
X Per\dotfill & 53290.922 &  0.740 &     $-$17.71 & 
\phn   $-9.7$ & \phn$ -80.7$ & \phn     53.6 & 0.66 & 0.87 & 2.65 & 3.46 \\
X Per\dotfill & 53290.926 &  0.740 &     $-$17.61 & 
\phn   $-9.4$ & \phn$ -80.7$ & \phn     56.8 & 0.81 & 0.92 & 2.89 & 3.28 \\
X Per\dotfill & 53292.922 &  0.748 &     $-$18.12 & 
\phn   $-9.3$ & \phn$ -78.3$ & \phn     56.2 & 0.78 & 0.93 & 2.74 & 3.26 \\
X Per\dotfill & 53292.928 &  0.748 &     $-$17.93 & 
\phn   $-9.1$ & \phn$ -82.4$ & \phn     51.7 & 0.59 & 0.83 & 2.57 & 3.64 \\
X Per\dotfill & 54019.950 &  0.653 &     $-$23.81 & 
\phn   $-6.4$ &\phn\phn$-8.9$& \nodata     &\nodata&\nodata& 7.92 &\nodata \\
X Per\dotfill & 54020.939 &  0.657 &     $-$23.61 & 
\phn   $-4.4$ &\phn\phn$-6.9$& \nodata     &\nodata&\nodata& 7.78 &\nodata \\
X Per\dotfill & 54021.930 &  0.661 &     $-$24.38 & 
\phn   $-4.5$ &\phn\phn$-7.2$& \nodata     &\nodata&\nodata& 7.78 &\nodata \\
\enddata
\end{deluxetable}

\clearpage


\begin{deluxetable}{lcc}
\tabletypesize{\scriptsize}
\tablewidth{0pt}
\tablecaption{Adopted Stellar and Disk Parameters\label{tab2}}
\tablehead{
\colhead{Parameter} &
\colhead{HDE 245770} &
\colhead{X Per} 
}
\startdata
 $R_s$ ($R_\odot$)       \dotfill & 15.   &  6.5  \\
 $M_s$ ($M_\odot$)       \dotfill & 20.   & 15.5  \\
 $T_s$ (K)               \dotfill & 28000 & 29500 \\
 $i$ (deg)               \dotfill & 28.5  & 31.5  \\
 $R_{\rm outer}$ ($R_s$) \dotfill & 15.9  & 42.9  \\
\enddata
\end{deluxetable}



\begin{deluxetable}{lcc}
\tabletypesize{\scriptsize}
\tablewidth{0pt}
\tablecaption{{\it IUE} Radial Velocity Measurements\label{tab3}}
\tablehead{
\colhead{Date} &
\colhead{Orbital} &
\colhead{$V_r$} 
\\
\colhead{(HJD$-$2,400,000)} &
\colhead{Phase} &
\colhead{(km s$^{-1}$)} 
}
\startdata
 43602.749\dotfill & 0.034 & \phn  $-$5.3 \\
 43705.644\dotfill & 0.445 & \phn\phs 3.3 \\
 43712.142\dotfill & 0.471 &     \phs20.4 \\
 43784.061\dotfill & 0.758 & \phn\phs 6.2 \\
 43796.362\dotfill & 0.808 & \phn\phs 7.5 \\
 43850.204\dotfill & 0.023 & \phn\phs 4.3 \\
 43885.683\dotfill & 0.164 & \phn\phs 5.8 \\
 43885.749\dotfill & 0.165 & \phn  $-$3.5 \\
 43885.806\dotfill & 0.165 & \phn  $-$5.6 \\
 43947.719\dotfill & 0.412 & \phn\phs 1.6 \\
 44156.035\dotfill & 0.245 & \phn  $-$0.7 \\
 44231.213\dotfill & 0.545 & \phn\phs 3.8 \\
 44319.029\dotfill & 0.896 & \phn  $-$3.2 \\
 44319.087\dotfill & 0.896 & \phn  $-$5.2 \\
 44319.138\dotfill & 0.896 & \phn  $-$0.3 \\
 44319.187\dotfill & 0.896 & \phn  $-$4.3 \\
 44319.238\dotfill & 0.897 & \phn  $-$4.3 \\
 44319.279\dotfill & 0.897 & \phn  $-$9.4 \\
 44323.732\dotfill & 0.915 & \phn\phs 2.3 \\
 44327.924\dotfill & 0.931 & \phn\phs 0.7 \\
 44582.952\dotfill & 0.950 & \phn  $-$0.6 \\
 44582.991\dotfill & 0.950 & \phn\phs 4.2 \\
 44888.132\dotfill & 0.169 & \phn  $-$3.9 \\
 48118.467\dotfill & 0.075 & \phn  $-$4.4 \\
 49593.258\dotfill & 0.967 & \phn\phs 5.8 \\
 49593.325\dotfill & 0.968 & \phn\phs 2.1 \\
 49621.183\dotfill & 0.079 & \phn  $-$0.2 \\
 49621.275\dotfill & 0.079 & \phn\phs 1.1 \\
 49643.343\dotfill & 0.168 & \phn  $-$4.0 \\
 49647.365\dotfill & 0.184 & \phn  $-$1.0 \\
 49715.074\dotfill & 0.454 & \phn\phs 2.5 \\
 49715.112\dotfill & 0.454 &     \phs10.0 \\
 49715.159\dotfill & 0.454 & \phn\phs 6.3 \\
 49715.198\dotfill & 0.455 & \phn\phs 0.2 \\
 49748.731\dotfill & 0.589 & \phn\phs 0.2 \\
 49748.796\dotfill & 0.589 & \phn  $-$0.9 \\
 49775.778\dotfill & 0.697 & \phn\phs 8.2 \\
 49775.824\dotfill & 0.697 & \phn\phs 7.0 \\
 49956.299\dotfill & 0.418 & \phn\phs 2.9 \\
 50000.094\dotfill & 0.593 & \phn  $-$4.1 \\
 50104.187\dotfill & 0.009 &      $-$15.1 \\
 50124.212\dotfill & 0.089 & \phn\phs 1.2 \\
 50144.055\dotfill & 0.168 & \phn\phs 1.0 \\
\enddata
\end{deluxetable}



\clearpage

\clearpage

\clearpage

\input{epsf}

\clearpage

\clearpage

\clearpage

\end{document}